\documentclass[12pt]{article}
\usepackage{amsfonts,amssymb}

\def\Z{\mathbb{Z}}

\def\defeq {\stackrel{\mbox{\rm\small def}}{=}}

\def\bq{ \begin{equation} }
\def\eq{ \end{equation} }
\def\ben{ \begin{eqnarray} }
\def\en{ \end{eqnarray} }

\def\frac#1#2{{#1\over #2}}

\def\on#1#2{\mathop{\vbox{\ialign{##\crcr\noalign{\kern2pt}
$\scriptstyle{#2}$\crcr\noalign{\kern2pt\nointerlineskip}
\kern-2pt$\hfil\displaystyle{#1}\hfil$\crcr}}}\limits}

\begin{document}

\baselineskip=15pt
\vspace{1cm} \centerline{{\LARGE \textbf {Classification of integrable hydrodynamic chains
 }}}

\vskip1cm \hfill
\begin{minipage}{13.5cm}
\baselineskip=15pt {\bf A.V. Odesskii ${}^{1,}{}^{2}$,  V.V. Sokolov
${}^{1}$}
\\ [2ex] {\footnotesize
${}^{1}$  L.D. Landau Institute for Theoretical Physics (Russia)
\\
${}^{2}$  Brock University (Canada)
\\}
\vskip1cm{\bf Abstract} Using the method of hydrodynamic reductions, we find all integrable infinite (1+1)-dimensional  hydrodynamic-type chains of shift one. A class of integrable infinite (2+1)-dimensional hydrodynamic-type chains is constructed.

\end{minipage}

\vglue1cm \textbf{Address}: L.D. Landau Institute for Theoretical
Physics of Russian Academy of Sciences, Kosygina 2, 119334,
Moscow, Russia

\textbf{E-mail}: aodesski@brocku.ca,  sokolov@itp.ac.ru
\newpage
\tableofcontents
\newpage
\section{Introduction}

We consider integrable infinite quasilinear chains of the form
\begin{equation}   \label{genchain}
u_{\alpha, t}=\phi_{\alpha,1} u_{1,x}+\cdots+\phi_{\alpha,\alpha+1} u_{\alpha+1, x}, \qquad \alpha=1,2,..., \qquad \phi_{\alpha,\alpha+1}\ne 0,
\end{equation}
where $\phi_{\alpha,j}=\phi_{\alpha,j}(u_1,...,u_{\alpha+1}).$ Two chains are called {\it equivalent} if they are
related by a transformation of the form
\begin{equation}   \label{tran}
u_{\alpha} \rightarrow \Psi_{\alpha} (u_1,...,u_{\alpha}),  \qquad
\frac{\partial \Psi_{\alpha}}{\partial u_{\alpha}}\ne 0, \qquad \alpha=1,2,...
\end{equation}

By integrability we mean the existence of an infinite set of hydrodynamic reductions \cite{ferhus1,ferhus2,ferhus3,pav4,pav1,fermar}.

{\bf Example 1.} The Benney equations \cite{benney, kupman, zakh}
\begin{equation}   \label{benny}
u_{1, t}=u_{2, x}, \qquad u_{2, t}=u_1 u_{1,x}+u_{3, x}, \dots \quad u_{\alpha t}=(\alpha-1) u_{\alpha-1} u_{1, x}+u_{\alpha+1, x}, \dots
\end{equation}
provide the most known example of integrable  chain (\ref{genchain}). The hydrodynamic reductions for the Benney chain were investigated in \cite{Gibt}. $\square$

In \cite{pav4,pav1,fermar} integrable divergent chains of the form
\begin{equation} \label{chain1}
u_{1t}=F_1(u_1,u_2)_x, \quad u_{2t}=F_2(u_1,u_2,u_3)_x, \cdots , \quad u_{it}=F_i(u_1,u_2,...,u_{i+1})_x,\cdots
\end{equation}
were considered.  In \cite{fermar} some necessary integrability conditions were obtained. Namely, a non-linear overdetermined system of PDEs for functions $F_1, F_2$ was presented. The general solution of the system was not
found. Another open problem was to prove that the conditions are sufficient. In other words, for any solution $F_1,F_2$ of the system  one should find functions
$F_i, i>2$ such that the resulting chain is integrable.

Probably any integrable chain (\ref{genchain}) is equivalent to a divergent chain. However, the divergent coordinates
are not suitable for explicit formulas. Our main observation is that a convenient coordinates are those, in which the
so-called Gibbons-Tsarev type system (GT-system) related to integrable chain is in a canonical form.

Using our version (see \cite{odsok4,odsok5}) of the hydrodynamic reduction method, we describe all integrable chains
(\ref{genchain}).  We establish an one-to-one correspondence between integrable chains  (\ref{genchain}) and  infinite triangular GT-systems
of the form
\begin{equation}\label{gt11}
\partial_i p_j=\frac{P(p_i,p_j)}{p_i-p_j}\,\partial_i u_1, \qquad ~i\ne
j,~
\end{equation}
\begin{equation}\label{gt12}
\partial_{i} \partial_{j}u_1=\frac{Q(p_i,p_j)}{(p_i-p_j)^2}\, \partial_i u_1\partial_j u_1,\qquad ~i\ne
j,~
\end{equation}
$$
\partial_i u_m=(g_{m,0}+g_{m,1} p_i+\cdots+g_{m,m-1} p_i^{m-1})\,\partial_i u_1, \qquad g_{m,j}=
g_{m,j}(u_1,...,u_{m}),\quad g_{m,m-1}\ne 0,
$$
where $m=2,3,...$ and $i,j=1,2,3.$  The functions $P,Q$ are polynomials quadratic in each of variables $p_i$ and
$p_j,$ with coefficients being functions of $u_1,u_2.$
The functions $p_1,p_2,p_3$, $u_1,u_2,...$ in (\ref{gt}) depend on $r^1,r^2,r^3,$  and $\partial_i=\frac{\partial}{\partial r^i}$.

{\bf Example 1-1} (continuation of Example 1.) The system (\ref{gt11}),(\ref{gt12})  corresponding to the Benney chain has the following form
\begin{equation}\label{gtBenni}
\partial_ip_j=\frac{\partial_i u_1}{p_i-p_j},  \qquad
\partial_{i} \partial_{j} u_1=\frac{2\, \partial_i u_1  \, \partial_j u_1}{(p_i-p_j)^2},\end{equation}
\begin{equation}\label{gtBenni1}
\partial_i u_m=(-(m-2) u_{m-2}-\cdots-2 u_2 p_i^{m-2}-u_1 p_i^{m-3}+ p_i^{m-1})\,\partial_i u_1.
\end{equation}
Equations (\ref{gtBenni}) were firstly obtained in \cite{Gibt}. $\square$

Given GT-system (\ref{gt11}), (\ref{gt12})  the coefficients of (\ref{genchain}) are uniquely defined by the following
relations
\begin{equation}   \label{relat}
p_i \partial_i u_m=\phi_{m,1} \partial_i u_1+\cdots+\phi_{m,m+1} \partial_i u_{m+1}, \qquad m=2,3,...
\end{equation}
Namely, equating the coefficients at different powers of $p_i$ in (\ref{relat}), we get a triangular system of linear
algebraic equations for $\phi_{i,j}$. Thus, the classification problem for chains (\ref{genchain}) is reduced to a description of all GT-systems (\ref{gt11}), (\ref{gt12}) . The latter problem is solved in Section 4-6.

The paper is organized as follows.  Following \cite{odsok4, odsok5}, we recall main definitions in Section 2 (see \cite{ferhus1,ferhus2,ferhus3,odsok4} for details). We consider only 3-component hydrodynamic reductions since the
existence of reductions with  $N>3$   gives nothing new \cite{ferhus1}. In Section 3 we formulate our previous results that are needed in the paper. Section 4 is devoted to a classification of admissible polynomials $P$ and $Q$ in (\ref{gt11}), (\ref{gt12}). In Sections 5,6 we construct integrable  chains for the generic case and for some degenerations. Section 6 also contains  examples of (2+1)-dimensional infinite hydrodynamic-type chains integrable from the viewpoint of the method of hydrodynamic reductions. Infinitesimal symmetries of GT-systems are studied in Section 7. These symmetries seem to be   important basic objects in the hydrodynamic reduction approach. 

\vskip.3cm \noindent {\bf Acknowledgments.} Authors thank M.V. Pavlov for fruitful
discussions.   V.S. is grateful to Brock University for hospitality. He was partially supported by
the RFBR grants 08-01-464, 09-01-22442-KE, and NS 3472.2008.2.

\section{Integrable chains and hydrodynamic reductions}

According to \cite{ferhus1,ferhus2,ferhus3,pav4,pav1,fermar} a chain (\ref{genchain}) is called {\it integrable} if it admits sufficiently many so-called hydrodynamic reductions.

{\bf Definition.} A hydrodynamic (1+1)-dimensional $N$-component reduction of a chain (\ref{genchain}) is a semi-Hamiltonian (see formula (\ref{semiham}) ) system of the form
\begin{equation}\label{red}r^i_t=p_i(r^1,...,r^N)r^i_x, \qquad i=1,.., N \end{equation}
and functions $u_j(r^1,...,r^N),~j=1,2,...$ such that for each solution of (\ref{red}) functions $u_j=u_j(r^1,...,r^N),~i=1,...$ satisfy (\ref{genchain}).

Substituting $u_i=u_i(r^1,...,r^N),~i=1,...$ into (\ref{genchain}), calculating $t$ and $x$-derivatives by virtue of (\ref{red}) and equating coefficients at $r^s_x$ to zero, we obtain
$$\partial_{s}u_{\alpha}p_s=\phi_{\alpha,1} \partial_{s}u_{1}+\cdots+\phi_{\alpha,\alpha+1} \partial_{s}u_{\alpha+1}, \qquad \alpha=1,2,...$$
It is clear from this system that
$$
\partial_su_k=g_k(p_s,u_1,...,u_k)\,\partial_s u_1,\qquad ~k=2,3,...
$$
where $g_k(p,u_1,...,u_k)$ is a polynomial of degree $k-1$ in $p$ for each $k=2,3,...$ Compatibility conditions $\partial_i\partial_ju_k=\partial_j\partial_iu_k$ give us a system of linear equations for $\partial_ip_j,~\partial_jp_i,~\partial_i\partial_ju_1,~i\ne j$. This system should have a solution (otherwise we would not have sufficiently many reductions). Moreover, expressions for $\partial_su_k,~k=2,3,...,~\partial_jp_i,~\partial_i\partial_ju_1,~i\ne j$ should be compatible and form a so-called GT-system.

{\bf Remark.} In the sequel we assume $N=3$ because the case $N>3$ gives nothing new \cite{ferhus1}.

\section{GT-systems}

{\bf Definition.} A compatible system of PDEs of the form
$$
\partial_i p_j=f(p_i,p_j,u_1,...,u_n),\,\partial_i u_1 \qquad ~j\ne
i,
$$
\begin{equation}\label{gt}
\partial_{i} \partial_{j}u_1=h(p_i,p_j,u_1,...,u_n)\, \partial_i u_1\partial_j u_1,\qquad ~j\ne
i,
\end{equation}
$$
\partial_iu_k=g_k(p_i,u_1,...,u_n)\,\partial_i u_1,\qquad ~k=1,...,n-1,
$$
where $~i,j=1,2,3$ is called $n$-{\it fields GT-system}.
Here $p_1,p_2,p_3$, $u_1,...,u_n$ are functions of $r^1,r^2,r^3$   and $\partial_i=\frac{\partial}{\partial {r^i}}$.

{\bf Definition.} Two GT-systems are called {\it equivalent} if they are related by a transformation of the form
\begin{equation}\label{gr}
p_i\to\lambda(p_i,u_1,...,u_n),\qquad
\end{equation}
\begin{equation}\label{gr1}
u_k\to\mu_k(u_1,...,u_n),\qquad ~k=1,...,n.
\end{equation}

{\bf Example 2} \cite{odsok1}. Let $a_0,a_1,a_2$ be arbitrary constants, $R(x)=a_2 x^2+a_1 x+a_0$. Then the system
\begin{equation}\label{gibtsar1}
\partial_ip_j=\frac{a_2 p_j^2+a_1 p_j+a_0}{p_i-p_j}\partial_i u_{1}, \qquad \partial_{i} \partial_{j} u_1=\frac{2 a_2 p_i p_j+a_1(p_i+p_j)+2 a_0}{(p_i-p_j)^2}\partial_i u_{1}\partial_j u_{1}
 \end{equation}
is an one-field GT-system. The original Gibbons-Tsarev system (\ref{gtBenni}) corresponds to
$a_2=a_1=0, a_0=1.$ The polynomial $R(x)$ can be reduced to one of the following canonical forms: $R=1$, $R=x$, $R=x^2$, or $R=x (x-1)$ by a linear transformation (\ref{gr}). A wide class of integrable 3D-systems of hydrodynamic type
 related to (\ref{gibtsar1}) is described in \cite{odsok1}. An
elliptic version of this GT-system and the corresponding integrable 3D-systems were constructed in \cite{odsok2}.
$\square$

{\bf Definition.} An additional system
\begin{equation}\label{gtex}
\partial_iu_k=g_k(p_i,u_1,...,u_{n+m})\partial_iu_n, \qquad k=n+1,...,n+m
\end{equation}
such that (\ref{gt}) and (\ref{gtex}) are compatible is called {\it an extension} of (\ref{gt}) by fields
$u_{n+1},...,u_{n+m}.$

It turns our that
$$
\partial_i u_{n+1}=f(p_i,u_{n+1},u_1,...,u_n) \,\partial_i u_{1}
$$
is an extension for GT-system (\ref{gt}). Stress that here $f$ is the same function as in (\ref{gt}). We call this extension  {\it the regular extension} by $u_{n+1}$.

{\bf Example 2-1.} The generic case of Example 2 corresponds to $R=x(x-1)$. The regular extension by $u_2$ is given by
$$
\partial_{i} u_2=\frac{u_2 (u_2-1)}{p_i-u_2}\partial_i u_1.
$$
If we express $u_1$ from this formula and substitute it to (\ref{gibtsar1}), we get the following one-field GT-system
\begin{equation} \begin{array}{c}
\displaystyle \partial_ip_j=\frac{p_j (p_j-1) (p_i-u_1)}{u_1 (u_1-1)(p_i-p_j)}\partial_i u_1, \\[7mm]
\displaystyle  \partial_{i} \partial_{j} u_1=\frac{p_i p_j (p_i+p_j)-p_i^2-p_j^2 + (p_i^2+p_j^2-4 p_i p_j+p_i+p_j) u_1}{u_1 (u_1-1)(p_i-p_j)^2}\partial_i u_{1}\partial_j u_{1}. \qquad \square
\end{array} \label{GTu}
\end{equation}

The second basic notion of the hydrodynamic reduction method is so-called  GT-family of (1+1)-dimensional hydrodynamic-type systems.

{\bf Definition.} An (1+1)-dimensional 3-component
hydrodynamic-type system of the form
\begin{equation}\label{gtgen}
r^i_t=v^i(r^1,...,r^N)\, r^i_x, \qquad i=1,2,3,
\end{equation}
is called semi-Hamiltonian if the following relation holds
\begin{equation}   \label{semiham}
\partial_{j}\frac{\partial_{i} v^{k}}{v^{i}-v^{k}}=\partial_{i}\frac{\partial_{j}
v^{k}}{v^{j}-v^{k}}, \qquad  i\ne j\ne k.
\end{equation}

{\bf Definition.} A Gibbons-Tsarev family associated with the
Gibbons-Tsarev type system (\ref{gttr}) is a (1+1)-dimensional
hydrodynamic-type system of the form
\begin{equation}\label{gtf}
r^i_t=F(p_i,u_1,...,u_m)r^i_x, \qquad i=1,2,3,
\end{equation}
semi-Hamiltonian by virtue of (\ref{gt}).

{\bf Example 2-2} \cite{odsok1}. Applying the regular extension to the generic GT-system (\ref{gibtsar1}) two times,
we get the following GT-system:
 \begin{equation}\label{gibtsar2}
\partial_ip_j=\frac{p_j(p_j-1)}{p_i-p_j}\partial_i w, \qquad
\partial_{ij}w=\frac{2p_ip_j-p_i-p_j}{(p_i-p_j)^2}\partial_iw\partial_jw,\qquad i\ne
j,
\end{equation}
 \begin{equation}\label{u1}
\partial_iu_j=\frac{u_j(u_j-1)\partial_iw}{p_i-u_j}, \qquad  j=1,2.
\end{equation}
Consider the generalized hypergeometric \cite{gel} linear system of the form
 \begin{equation}\label{hyp1}
\frac{\partial^2 h}{\partial u_j \partial
u_k}=\frac{s_j}{u_j-u_k}\cdot \frac{\partial h}{\partial
u_k}+\frac{s_k}{u_k-u_j}\cdot \frac{\partial h}{\partial
u_j},~ \quad j\ne k,
\end{equation}
\begin{equation}\label{hyp2} \begin{array}{c}   \displaystyle \frac{\partial^2
h}{\partial u_j \partial u_j}=-\left(1+\sum_{k=1}^{n+2} s_k\right)
\frac{s_j}{u_j (u_j-1) }\cdot
h+  \frac{s_j}{u_j (u_j-1)} \sum_{k\ne j}^n \frac{u_k
(u_k-1)}{u_k-u_j}\cdot
\frac{\partial h}{\partial u_k}+\\[10mm]
\displaystyle \left(\sum_{k\ne j}^n \frac{s_k}{u_j-u_k}+
\frac{s_j+s_{n+1}}{u_j}+ \frac{s_j+s_{n+2}}{u_j-1}\right)\cdot
\frac{\partial h}{\partial u_j}.
\end{array}
\end{equation}
Here $i,j=1,2$ and $s_1,...,s_{4}$ are arbitrary parameters. It easy to verify that this system is in involution and therefore the solution space is 3-dimensional.  Let $h_1,h_2,h_3$ be a basis of this space. For any $h$ we put
$$S(p,h)=u_1(u_1-1)(p-u_2)\frac{h h_{1,u_1}-h_{u_1}h_1}{h_1}+u_2(u_2-1)(p-u_1)\frac{h h_{1,u_2}-h_{u_2}h_1}{h_1}.$$
Then the formula
\begin{equation}\label{FF}
F=\frac{S(p,h_3)}{S(p,h_2)}
\end{equation}
defines the generic linear fractional GT-family for (\ref{gibtsar2}). $\square$

\section{Canonical forms of GT-systems associated \\ with integrable chains}

For integrable chains the corresponding GT-systems involve infinite number of fields $u_i,~i=1,2,...$ (see Example 1-1). In this Section we show that these GT-systems are equivalent to infinite triangular extensions of one-field GT-systems from Examples 2,3.

A compatible system of PDEs of the form
$$
\partial_i p_j=f(p_i,p_j,u_1,...,u_n)\,\partial_i u_1 \qquad ,~i\ne
j,
$$
\begin{equation}\label{gttr}
\partial_iu_k=g_k(p_i,u_1,...,u_k)\,\partial_i u_1,\qquad ~k=1,2,..., ,
\end{equation}
$$
\partial_{i} \partial_{j}u_1=h(p_i,p_j,u_1,...,u_n)\, \partial_i u_1\partial_j u_1,\qquad ~i\ne
j,
$$
where $~i,j=1,2,3$ is called  {\it triangular GT-system}.
Here $p_1,p_2,p_3$, $u_1,u_2,...$ are functions of $r^1,r^2,r^3,$ and $\partial_i=\frac{\partial}{\partial {r^i}}$.

{\bf Definition.} A chain (\ref{genchain}) is called integrable if
there exists a Gibbons-Tsarev type system of the form (\ref{gttr}) and
a Gibbons-Tsarev family
\begin{equation}\label{gtfcom}
r^i_t=F(p_i,u_1,...,u_m)r^i_x,\qquad i=1,2,3,
\end{equation}
such that (\ref{genchain}) holds by virtue of (\ref{gttr}),
(\ref{gtfcom}).

Due to the equivalence transformations (\ref{gr}) we can assume without loss of generality that
\begin{equation}\label{ff}
F(p,u_1,...,u_m)=p.
\end{equation}
Under this assumption we have
$$u_{j,t}=\sum_s\partial_{s}u_jr^s_t=\sum_s\partial_{s}u_jp_sr^s_x.$$
and similar
$$u_{j,x}=\sum_s\partial_{s}u_jr^s_x.$$
Substituting these expressions into (\ref{genchain}) and equating coefficients at $r^s_x$ to zero, we obtain
$$\partial_{s}u_{\alpha}p_s=\phi_{\alpha,1} \partial_{s}u_{1}+\cdots+\phi_{\alpha,\alpha+1} \partial_{s}u_{\alpha+1}, \qquad \alpha=1,2,...$$
Using (\ref{gttr}) and replacing $p_s$ by $p$, we get 
$$p=\phi_{1,1}+\phi_{1,2}g_2,~pg_2=\phi_{2,1}+\phi_{2,2}g_2+\phi_{2,3}g_3,~pg_3=\phi_{3,1}+\phi_{3,2}g_2+\phi_{3,3}g_3+\phi_{3,4}g_4,...$$
Solving this system with respect to $g_2,~g_3,...$, we obtain
$$g_i(p)=\psi_{i,0}+\psi_{i,1}p+...+\psi_{i,i-1}p^{i-1}.$$
Here $\psi_{i,j}$ are functions of $u_1,...,u_i$. For example,
\begin{equation}\label{gg}
g_2=-\frac{p}{\phi_{1,2}}-\frac{\phi_{1,1}}{\phi_{1,2}}.
\end{equation}

{\bf Remark.} Since we assume that $\phi_{i,i-1}\ne0,$ we have $\psi_{i,i-1}\ne0$ for all $i$. 
Therefore $g_1=1,g_2,...$ is a basis in the linear space of all polynomials in $p$.
The coefficients $\phi_{i,j}$ of our chain  are just entries of the matrix of multiplication by $p$ in this basis.
More generally, if we don't normalize $F=p$, then the coefficients $\phi_{i,j}$ can be found from the equations
\begin{equation}\label{coefchain} \begin{array}{l}
F(p)=\phi_{1,1}+\phi_{1,2}g_2,\qquad
F(p)g_2=\phi_{2,1}+\phi_{2,2}g_2+\phi_{2,3}g_3,\\[4mm]
F(p)g_3=\phi_{3,1}+\phi_{3,2 }g_2+\phi_{3,3}g_3+\phi_{3,4}g_4,...$$
\end{array}
\end{equation}

Compatibility conditions $\partial_i\partial_ju_{\alpha}=\partial_j\partial_iu_{\alpha},~\alpha=2,3,4$
give a system of linear equations for $\partial_ip_j,~\partial_jp_i,~\partial_i\partial_ju_1$.
Solving this system, we obtain formulas (\ref{gt11}),(\ref{gt12}), where in principal $P,~Q$ could
depend on $u_1,u_2,u_3,u_4$. However, it follows from compatibility conditions
$\partial_i\partial_jp_k=\partial_j\partial_ip_k$ that $P,~Q$ depend on $u_1,~u_2$ only.

Written (\ref{gt11}) in the form
\begin{equation}\label{gtnew1}
\partial_i p_j=\left( \frac{R(p_j)}{p_i-p_j}+(z_4 p_j^2+z_5 p_j+z_6) p_i+ z_4 p_j^3+z_3 p_j^2+z_7 p_j+z_8 \right)\,\partial_i u_1,
\end{equation}
where $R(x)=z_4 x^4+z_3 x^3+z_2 x^2+z_1 x+z_0,$ one can derive from the compatibility conditions $\partial_i\partial_jp_k=\partial_j\partial_ip_k$,  $\partial_i\partial_j u_1=\partial_j\partial_i u_1$
that the equation (\ref{gt12}) has the following form
\begin{equation}\label{gtnew2}
\partial_{i} \partial_{j}u_1=\left(\frac{2 z_4 p_i^2 p_j^2+z_3 p_i p_j (p_i+p_j)+z_2 (p_i^2+p_j^2)+z_1(p_i+p_j)+2 z_0 }{(p_i-p_j)^2}+z_9\right)\, \partial_i u_1\partial_j u_1.
\end{equation}

It is easy to verify that we can normalize $z_9=z_6-z_7,$ $g_2=p$ by a transformation (\ref{tran}). Then the coefficients $z_i(x,y), i=0,...,8$ satisfy the following pair of compatible dynamical systems with respect to  $y$ and  $x$:  
$$ \begin{array}{l} 
z_{0,y}=2 z_0 z_5-z_1 z_6,  \qquad z_{1,y}=4 z_0 z_4+z_1 z_5-2 z_2 z_6,  \qquad z_{2,y}=3 z_1 z_4-3 z_3 z_6 ,
\\[3mm] z_{3,y}=2 z_2 z_4-z_3 z_5-4 z_4 z_6,  \qquad z_{4,y}=z_3 z_4 -2 z_4 z_5,  \qquad z_{5,y}=z_4 z_7-z_4 z_6-z_5^2 ,
\\[3mm] z_{6,y}=z_4 z_8-z_5 z_6,  \qquad z_{7,y}=2 z_1 z_4-2 z_3 z_6 -z_5 z_6+z_4 z_8,  \qquad z_{8,y}=2 z_0 z_4-z_6^2-z_6 z_7+z_5 z_8,
\end{array} 
$$
and
$$ \begin{array}{l} 
z_{0,x}=-z_0 z_2-z_0 z_6+3 z_0 z_7-z_1 z_8,  \qquad z_{1,x}= -z_1 z_2+3 z_0 z_3-z_1 z_6+2 z_1 z_7-2 z_2 z_8,  \\[3mm] z_{2,x}=-z_2^2+2 z_1 z_3+4 z_0 z_4-z_2 z_6+z_2 z_7-3 z_3 z_8, \qquad z_{3,x}=3 z_1 z_4-z_3 z_6-4 z_4 z_8, 
\\[3mm] z_{4,x}=z_2 z_4 -z_4 z_6-z_4 z_7,  \qquad z_{5,x}=z_1 z_4-z_5 z_6-z_4 z_8 ,  \qquad z_{6,x}=z_0 z_4-z_6^2,
\\[3mm]   z_{7,x}=z_1 z_3+3 z_0 z_4+z_1 z_5-z_2 z_6-z_2 z_7+z_7^2-z_3 z_8-2 z_5z _8, \\[3mm] z_{8,x}=z_0 z_3+z_0 z_5-z_2 z_8-2 z_6 z_8+z_7 z_8.
\end{array} 
$$
These is a complete description of the GT-systems related to integrable chains  (\ref{genchain}). 
 
To solve the dynamical systems we bring the polynomial $R$
to a canonical form sacrificing to the normalization (\ref{ff}).   

It is obvious that linear transformations $p_i \rightarrow a p_i+b$, where $a,b$ are functions of $u_1,u_2,$ preserve the form of GT-system (\ref{gtnew1}),(\ref{gtnew2}). Moreover, there exist transformations of the form
\begin{equation}\label{tr}p_i = \frac{a \bar p_i+b}{\bar p_i-\psi}, \qquad i=1,2,3\end{equation}
preserving the form of GT-system (\ref{gtnew1}),(\ref{gtnew2}). Such admissible transformations are described by the following conditions:
$$
a_{u_{2}}= z_4 (b+a \psi), \qquad  b_{u_{2}}=z_4 b \psi+z_5 b-z_6 a,\qquad \psi_{u_{2}}=z_4 \psi^2+z_5 \psi+z_6.
$$

Under transformations (\ref{tr})  the polynomial $R$  is transformed by the following simple way:
$$
R(p_i) \rightarrow   (p_i-\psi)^4 R\Big(\frac{a p_i+b}{p_i-\psi}\Big).
$$
Suppose that $R$ has distinct roots. It is possible to verify that by an admissible transformation (\ref{tr}) we can move three of the four roots  to $0,~1$ and $\infty$. It follows from compatibility conditions for the GT-system that then the fourth root $\lambda(u_1,u_2)$ does not depend on $u_2$.
Making transformation of the form $u_1\to q(u_1)$ we arrive at the canonical forms  $\lambda=u_1$ or $\lambda=const$.  It is straightforwardly verified that in the first case equations (\ref{gtnew1}), (\ref{gtnew2}) coincides with (\ref{GTu}).
In the second case the GT-system does not exist.

In the case of multiple roots the polynomial $R(x)$ can be reduced to one of the following forms:  $R=0$, $R=1$, $R=x$, $R=x^2$, or $R=x (x-1).$ In all these cases equations (\ref{gtnew1}), (\ref{gtnew2}) coincides with the corresponding
equations from Example 2.

Thus, the following statement is valid:

{\bf Proposition 1.} There are 6 non-equivalent cases of GT-systems (\ref{gtnew1}), (\ref{gtnew2}). The canonical forms are:
$$
\begin{array}{l}
{\bf Case~  1}: \quad (\ref{GTu})\qquad  \mbox{\rm (generic case)}; \\[1mm]
{\bf Case~  2}: \quad (\ref{gibtsar1})\quad \mbox{\rm with} \quad  R(x)=x (x-1); \\[1mm]
{\bf Case~  3}: \quad (\ref{gibtsar1}) \quad \mbox{\rm with} \quad  R(x)=x^2; \\[1mm]
{\bf Case~  4}: \quad (\ref{gibtsar1}) \quad \mbox{\rm with} \quad  R(x)=x; \\[1mm]
{\bf Case~  5}:\quad (\ref{gibtsar1}) \quad \mbox{\rm with} \quad  R(x)=1. \\[1mm]
{\bf Case~  6}:\quad (\ref{gibtsar1}) \quad \mbox{\rm with} \quad  R(x)=0.  \quad \square  
\end{array}
$$

{\bf Remark.} Cases 2-6 can be obtained from Case 1 by appropriate limit procedures. For example, Case 2 corresponds to the limit 
$u_1\rightarrow \frac{u_1}{\varepsilon}, \, \varepsilon \rightarrow 0.$

It follows from (\ref{ff}), (\ref{gg}) that for any canonical form the functions $F$ and $g_2$ have the following structure:
\begin{equation}\label{fg}
g_2(p_i)=\frac{k_1 p_i+k_2}{k_3 p_i+k_4}, \qquad F(p_i)=\frac{f_1 p_i+f_2}{k_3 p_i+k_4},
\end{equation}
where the coefficients are functions of $u_1,u_2$.

{\bf Lemma 1.} For the Case 1 any function $g_2$ can be reduced by an appropriate transformation $\bar u_2=\sigma(u_1,u_2)$ to one of the following canonical forms:
$$
\begin{array}{l}
{\bf a_1}: \displaystyle \quad g_2(p)=\frac{u_2 (u_2-1) (p-u_1)}{u_1 (u_1-1) (p-u_2)} \qquad  \mbox{\rm (regular extension)} ; \\[5mm]
{\bf b_1}: \displaystyle \quad g_2(p)=\frac{1}{p-u_1}; \\[5mm]
{\bf c_1}: \displaystyle \quad g_2(p)=\frac{u_1^{-\lambda} (u_1-1)^{\lambda-1}}{p-\lambda} \qquad \lambda=1,0;\\[5mm]
{\bf d_1}: \displaystyle \quad g_2(p)=\frac{u_1-u_2 }{u_1(u_1-1)} p+\frac{u_2-1}{u_1-1}. \quad \square
 \end{array}
$$

The GT-system from the Case 1 possesses a discrete automorphism group $S_4$ interchanging the points $0,1, \infty, u_1$. The group is defined by generators
 $$\displaystyle
\sigma_1: u_1\to 1-u_1, \quad \displaystyle
p_i\to 1-p_i, \qquad \sigma_2: u_1\to \frac{u_1}{u_1-1}, \quad \displaystyle
p_i\to \frac{p_i}{p_i-1},$$
and
$$ \sigma_3: u_1\to 1-u_1, \quad \displaystyle
p_i\to \frac{(1-u_1) p_i}{p_i-u_1}.$$
Up to this group the cases $b_1,c_1,d_1$ are equivalent and one can take say the case $d_1$ for further consideration.
The case $a_1$ is invariant with respect to the group.

{\bf Remark.} The cases $b_1,~c_1,~d_1$ are degenerations of the case $a_1$. Namely, they can be obtained as appropriate limit $u_2\to u_1$, $u_2\to \lambda,$ $u_2\to \infty$ correspondingly.

All possible functions $g_2$ for Cases 2-5 are described in the following

{\bf Lemma 2.} For the GT-system (\ref{gibtsar1}) (excluding Case 6) any function $g_2$ can be reduced by an appropriate transformation $\bar u_2=\sigma(u_1,u_2)$ to one of the following canonical forms:
$$
\begin{array}{l}
{\bf a_2}: \displaystyle \quad g_2(p)=\frac{R(u_2)}{p-u_2} \qquad  \mbox{\rm (regular extension)} ; \\[5mm]
{\bf b_2}: \displaystyle \quad g_2(p)=\frac{1}{p-\lambda}, \quad  \mbox{\rm where} \quad R(\lambda)=0; \\[5mm]
{\bf c_2}: \displaystyle \quad g_2(p)=p-a_2 u_2 .
 \end{array}
$$
The discrete automorphism of the GT-system interchanges the roots of $R$ in the case $b_2$. $\square$

{\bf Lemma 3.} For the GT-system (\ref{gibtsar1}) with $R(x)=0$ (Case 6) any function $g_2$ can be reduced  to $g_2(p)=p$ by an appropriate transformation $\bar u_2=\sigma(u_1,u_2)$. Furthermore, the corresponding triangular GT-system has the form
\begin{equation}\label{trivGT}
\partial_i p_j=0,\qquad \partial_i\partial_j u_1=0,\qquad \partial_i u_k=p_i^{k-1}u_1, \quad k=2,3,... \quad 
\square 
\end{equation}

\section{Generic case}

The next step in the classification is to find all functions $F$ of the form (\ref{gg}) for each pair consisting of a GT-system from Proposition 1 and the corresponding $g_2$ from Lemmas 1-3. The semi-Hamiltonian condition (\ref{semiham})
yields a non-linear system of PDEs for the functions $f_1(u_1,u_2)$, $f_2(u_1,u_2).$ For each case this system can be reduced to the linear generalized hyper\-geometric system (\ref{hyp1}), (\ref{hyp2}) with a special set of parameters $s_1,s_2,s_3,s_4$ or to a degeneration of this system.

The general linear fractional GT-family for the generic case 1, $\bf a_1$  is given by (\ref{FF}). According to (\ref{fg}), the additional restriction is that the root of the denominator has to be equal $u_2.$ It is easy to verify that this is equivalent to $s_2=0, h_{1,u_{2}}=h_{2,u_{2}}=0$.
The latter means that $h_1(u_1),h_2(u_1)$ are linear independent solutions of the standard hypergeometric equation
\begin{equation}\label{hyper1}
u (u-1) \, h(u)''+ [s_1+s_3-(s_3+s_4+2 s_1)\,u]\, h(u)'+s_1 (s_1+s_3+s_4+1)\, h(u)=0.
\end{equation}
The function $h_3(u_1,u_2)$ is arbitrary solution of (\ref{hyp1}), (\ref{hyp2}) with $s_2=0$ linearly independent of $h_1(u_1),h_2(u_1)$. Without loss of generality we can choose
$$
h_3(u_1,u_2)=\int_0^{u_{2}}(t-u_1)^{s_1} t^{s_{3}}(t-1)^{s_{4}}dt.
$$
Formula (\ref{FF}) gives
\begin{equation}\label{FFF}
F(p,u_1,u_2)=\frac{f_1(u_1,u_2)\, p-f_2(u_1,u_2)}{p-u_2},
\end{equation}
where
$$
f_1=\frac{u_2 (u_2-1) h_1 h_{3,u_2}+u_1 (u_1-1) (h_1 h_{3,u_1}-h_3 h_1')}{u_1 (u_1-1) (h_1 h_2'-h_2 h_1')},
$$
$$
f_2=\frac{u_1 u_2 (u_2-1) h_1 h_{3,u_2}+u_2 u_1 (u_1-1) (h_1 h_{3,u_1}-h_3 h_1')}{u_1 (u_1-1) (h_1 h_2'-h_2 h_1')}.
$$
Notice that $h_1 h_2'-h_2 h_1'=const (u_1-1)^{s_1+s_4} u_1^{s_1+s_3}.$

For integer values of $s_1,s_3,s_4$ the hypergeometric system can be solved explicitly.  For example, if $s_1=s_3=s_4=0$, the
above formulas give rise to $F=g_2.$ If $s_4=-2-s_1-s_3$ then
$$
F=\frac{(u_2-u_1)^{s_1+1} u_2^{s_3+1} (u_2-1)^{-1-s_1-s_3}}{p-u_2};
$$
if $s_4=0,$ then
$$
F=\frac{(p-1)(u_2-u_1)^{s_1+1} u_2^{s_3+1} (u_1-1)^{-1-s_1}}{p-u_2}.  
$$

Now we are to find the functions $g_3,g_4,... $ in (\ref{gttr}). These functions are define up to arbitrary transformation  (\ref{tran}), where $\alpha=3,4,...$. In practice, one can look for functions $g_3,g_4,... $ linear in $u_i, i>2$ (cf. (\ref{gtBenni1})). An extension linear in $u_i, i>2$ is given by
$$
g_3(p)=-\frac{(u_1-u_2)(u_2-1) p}{u_1 (u_1-1) (p-u_2)^2},
$$
$$
g_i(p)=\frac{(i-3)(u_1-u_2)(u_2-1) p\, u_i}{u_1 (u_1-1) (p-u_2)^2}-\frac{(u_1-u_2)^{i-3}(u_2-1)^2 p (p-u_1) (p-1)^{i-4}}{u_1 (u_1-1)^{i-2} (p-u_2)^{i-1}}\, -
$$
$$
\sum_{s=1}^{i-4} \frac{(i-s-2)(u_1-u_2)^{s}(u_2-1)^2 p (p-u_1) (p-1)^{s-1}\,u_{i-s}}{u_1 (u_1-1)^{s+1} (p-u_2)^{s+2}}.
$$

The coefficients of the  chain (\ref{genchain}) corresponding to Case 1, ${\bf a_1}$ are determined from  (\ref{coefchain}), where $F$ is given by (\ref{FFF}). Relations (\ref{coefchain}) are equivalent to a triangular system of linear algebraic equations. Solving this system, we find that for $i>4$ coefficients of the chain read:
$$
\phi_{i,i+1}=\frac{(u_1-1)(f_1 u_2-f_2)}{(u_2-1)(u_1-u_2)}\defeq Q_1, \qquad \phi_{i,i}=\frac{f_2-f_1}{u_2-1}\defeq Q_2,
$$
$$
\phi_{i,4}=-u_i Q_1 , \qquad \phi_{i,3}=-\Big( (u_4+i-3) u_i+(2-i) u_{i+1} \Big) Q_1 \defeq A_i,
$$
and $\phi_{i,j}=0$ for all remaining $i,j.$ For $i\leq 4$ we have 
\begin{equation}  \begin{array}{c} \label{ch1}
\displaystyle \phi_{1,1}=\frac{f_1 u_1-f_2}{u_1-u_2}, \qquad \phi_{1,2}=-\frac{u_1 }{u_2 } Q_1,
\\[5mm]
\displaystyle \phi_{2,1}=\frac{(u_2-1)(f_1 u_2-f_2)}{(u_1-1)(u_1-u_2)}, \qquad \phi_{2,2}=\frac{f_2 u_1-f_1 u_2^2}{u_2(u_1-u_2)}, \qquad \phi_{2,3}=f_1 u_2-f_2, 
\\[5mm]
\displaystyle \phi_{3,1}=\phi_{3,2}=0, \qquad \phi_{3,3}=Q_2-(u_4-1) Q_1, \qquad \phi_{3,4}=-Q_1,
\\[5mm]
\displaystyle \phi_{4,1}=\phi_{4,2}=0, \qquad \phi_{4,3}=A_4, \qquad \phi_{4,4}=Q_2-u_4 Q_1, \qquad \phi_{4,5}=Q_1.  
\end{array}
\end{equation}

The explicit formulas for other cases of Proposition 1 can be obtained by limits  from the above formulas. We outline the limit procedures for the case 1, $\bf d_1$.  In this case   the limit is given by $u_2\rightarrow u_1+\varepsilon u_{2},\quad \varepsilon\rightarrow 0 .$ It is easy to check that under this limit the extension $a_1$ turns to $d_1$. The limit of the system (\ref{hyp1}), (\ref{hyp2}) with $s_2=0$ can be easily found. The general solution of the system thus obtained is given by
$h=c_1 (u_2-u_1)^{1+s_1+s_3+s_4}+h_1,$ where $h_1$ is the general solution of (\ref{hyper1}).   Let $h_1,h_2$ be solutions of (\ref{hyper1}), and $h_3=(u_2-u_1)^{1+s_1+s_3+s_4}$. Then the limit procedure in (\ref{FFF}) gives rise to
$$
F(p,u_1,u_2)=Q\times \Big((1+s_1+s_3+s_4) h_1 (p-u_1)+u_1 (u_1-1) h_1'\Big),
$$
where
$$
Q=(u_2-u_1)^{1+s_1+s_3+s_4} (u_1-1)^{-1-s_1-s_4} u_1^{-1-s_1-s_3}.  
$$

As usual, the most degenerate cases in classification of integrable PDEs could be interesting for applications. In our classification they are  Case 5, $c_2$ and Case 6. 
The Benney chain (see Examples 1 and 1-1) belongs to  Case 5, case $c_2$ (i.e $g_2=p$).  Any GT-family has the form
$F=f_1(u_1,u_2) p+ f_2(u_1,u_2)$. If $f_1=1$ then $F=p+k_2 u_2+k_1 u_1.$ The Benney case corresponds to $k_1=k_2=0$. For arbitrary $k_i$
we get the Kupershmidt chain \cite{kup}. In the case $f_1=A(u_1), A'\ne 0$ we obtain:
$$
f_1=k_2 \exp(\lambda u_1)+k_1, \qquad f_2=k_2 k_3\exp(\lambda u_1)+\lambda k_1 (k_3 u_1-u_2).
$$
In the generic case
$$
F=\exp(\lambda u_2) (S_1(u_1) p+S_2(u_1)),
$$
where the functions $S_i$ can be expressed in terms of the Airy functions.

\section{Trivial GT-system and $2+1$-dimensional integrable hydrodynamic chains}

It was observed in \cite{odsok4} that (2+1)-dimensional systems of hydrodynamic type with the trivial GT-system usually admit some integrable multi-dimensional generalizations. For the chains  such GT-system  is defined by (\ref{trivGT}). That is why the Case 6 is of a great importance in our classification. 
The automorphisms 
of (\ref{trivGT}) are given by 
\begin{equation} \label{eqv1}
p_j\to p_j,~j=1,...,N,~u_i\to \nu u_i+\gamma_i,~i=1,2,...;
\end{equation}
$$p_j\to ap_j+b,~j=1,...,N,~u_i\to a^{i-1}u_i+(i-1)a^{i-2}bu_{i-2}+...+b^{i-1}u_1, \quad i=1,2,...$$

The corresponding GT-families are of the form $F(p)=A(u_1,u_2) p+B(u_1,u_2)$, where $A(x,y),B(x,y)$ satisfies the following system of PDEs:
\begin{equation}\begin{array}{c} \label{sys}
A B_{yy}=A_y B_y, \quad A B_{xy}=A_y B_x, \quad A B_{xx}=A_x B_x, \\[3mm]
A A_{yy}=A_y^2, \quad A A_{xy}=A_x A_y, \quad A A_{xx}=A_x^2+ A_x B_y-A_y B_x.
\end{array}
\end{equation}
This system can be easily solved in elementary functions. For each solution formula (\ref{coefchain}) defines the corresponding integrable chain (\ref{genchain}).

It follows from (\ref{sys}) that there are two types of $u_2$-dependence:

{\bf 1 } (generic case).\qquad  $F(p)=\exp(\lambda u_2)\Big(a(u_1) p+b(u_1)\Big)$,

{\bf 2.} \qquad \qquad \qquad \qquad $F(p)=a(u_1)p+\lambda u_2+b(u_1).$ 

In the first case there are two subcases: $b'\ne 0$ and $b'= 0.$ The first subcase gives rise to 
$$a=\sigma', \quad b=k_1 \sigma  \qquad \sigma(x)=c_1 \exp{(\mu_1 x)}+c_2 \exp{(\mu_2 x)},\quad \mbox{\rm where } \quad c_1 c_2 (\lambda k_1-\mu_1 \mu_2)=0.$$ 
The second subcase leads to 
$$
b=c_1, \qquad a(x)=c_2 \exp{(\mu x)}+c_3,\quad \mbox{\rm where } \quad c_2 (c_1 \lambda-c_3 \mu)=0.
$$

The same subcases for the case 2 yield 
$$a=\sigma', \quad b=k_1 \sigma  \qquad \sigma(x)=c_1+c_2 x+c_3 \exp{(\mu x)},\quad \mbox{\rm where } \quad  c_3 (\lambda -c_2 \mu)=0,$$ and 
$$
b=c_1, \qquad a(x)=c_2 \exp{(\mu x)}+c_3,\quad \mbox{\rm where } \quad c_2 (\lambda-c_3 \mu)=0.
$$

It is easy to verify that in the generic case the function $F$ can be reduced by ({\ref{eqv1}) to the form
$$F(p)=e^{u_2+u_1}(p-1)+e^{u_2-u_1}(p+1).$$
In this case the corresponding chain reads as
\begin{equation} \label{gg1}
u_{k,t}=(e^{u_2+u_1}+e^{u_2-u_1})u_{k+1,x}+(e^{u_2-u_1}-e^{u_2+u_1})u_{k,x}, \qquad k=1,2,3,...
\end{equation}
As usual, this chain is the first member of an infinite hierarchy. The second flow of this hierarchy is given by
$$u_{k,\tau}=(e^{u_2+u_1}+e^{u_2-u_1})u_{k+2,x}+(u_3-u_1)(e^{u_2+u_1}+e^{u_2-u_1})u_{k+1,x}+$$
$$(e^{u_2+u_1}(u_1-u_3-1)+e^{u_2-u_1}(u_3-u_1-1))u_{k,x}, \qquad k=1,2,3,...$$

In the case 2 with $c_3=\lambda=0, k_1=1$ we get the chain 
\begin{equation}  \label{alsh}
u_{k,t}= u_{k+1,x}+u_1 u_{k,x}, \qquad k=1,2,3,...
\end{equation}
This chain is equivalent to the chain of the so-called universal hierarchy \cite{alsh}.  The chain (\ref{alsh}) is a degeneration of the chain 
\begin{equation}  \label{alsh1}
u_{k,t}= u_{k+1,x}+u_2 u_{k,x}, \qquad k=1,2,3,...
\end{equation}

Following the line of \cite{ferhus3, odsok4} it is not difficult to find (2+1)-dimensional integrable generalizations for all (1+1)-dimensional integrable chains constructed above. 
Some families of functions $F$ described above linearly depend on two parameters. Denote these parameters by $\gamma_1,\gamma_2.$ 
The corresponding integrable chain 
$$
u_{k, t}=\gamma_1 (\phi_{k,1} u_{1,x}+\cdots+\phi_{k,k+1} u_{k+1, x})+\gamma_2 (\psi_{k,1} u_{1,x}+\cdots+\psi_{k,k+1} u_{k+1, x}) 
$$
is also linear in $\gamma_1,\gamma_2.$ We claim that the following (2+1)-dimensional chain 
\begin{equation}\label{3D}
u_{k, t}=(\phi_{k,1} u_{1,x}+\cdots+\phi_{k,k+1} u_{k+1, x})+(\psi_{k,1} u_{1,y}+\cdots+\psi_{k,k+1} u_{k+1, y}) 
\end{equation}
is integrable from the viewpoint of the method of hydrodynamic reductions. For each case the reductions can be easily   described.

For example, in the generic case 
$$F(p)=\gamma_1 e^{u_2+u_1}(p-1)+\gamma_2 e^{u_2-u_1}(p+1)$$
formula (\ref{3D}) yields (2+1)-dimensional chain  
\begin{equation}\label{3Dgen}
u_{k,t}=e^{u_2+u_1}(u_{k+1,x}-u_{k,x})+e^{u_2-u_1}(u_{k+1,y}+u_{k,y}), \qquad k=1,2,3,...
\end{equation}
After a change of variables of the form
$$x\to-\frac{1}{2}x, \quad y\to\frac{1}{2}y,\qquad u_1\to\frac{1}{2}u_0,\quad u_2\to u_1+\frac{1}{2}u_0,\quad u_3\to -2u_2+\frac{1}{2}u_0,...$$ (\ref{3Dgen}) can be written as
\begin{equation}\label{3Dgen1}
u_{0,t}=e^{u_1}u_{0,y}+e^{u_1}(u_{1,y}-e^{u_0}u_{1,x}),\qquad 
u_{i,t}=e^{u_0+u_1}u_{i,x}+e^{u_1}(e^{u_0}u_{i+1,x}-u_{i+1,y}),
\end{equation}
where $i=1,2,...$.
Probably (\ref{3Dgen1}) is a first example of a (2+1)-dimensional chain integrable from the viewpoint of the hydrodynamic reduction approach.  

Triangular GT-systems related to integrable  (2+1)-dimensional chains with fields $u_0,u_1,u_2,...$ have the form 
$$
\partial_ip_j=f_1(p_i, q_i,p_j, q_j,u_0,...,u_n)\partial_iu_0, \qquad \partial_iq_j=f_2(p_i, q_i,p_j, q_j,u_0,...,u_n)\partial_iu_0, 
$$
\begin{equation}\label{gt4d}
\partial_i\partial_j u_0=h(p_i,q_i,p_j,q_j,u_0,...,u_n)\partial_iu_0\partial_ju_0,  
\end{equation}
$$
\partial_iu_k=g_k(p_i,q_i,u_0,...,u_{k+1})\partial_iu_0, \qquad k=0,1,2,...  
$$
Here $i\ne j,~i,j=1,...,3$, $p_1,...,p_3,~q_1,...,q_3$, $u_0,u_1,u_2,..., $ are functions of $r^1,r^2,r^3.$ In particular, the GT-system associated with (\ref{3Dgen1})  has the form:
$$\partial_ip_j=\partial_i\partial_ju_0=0, \qquad \partial_iq_j=\Big(\frac{p_iq_i-p_jq_j}{p_i-p_j}-q_iq_j\Big)\partial_iu_0,\qquad \partial_iu_k=-\frac{p_i}{(p_i-1)^k}\partial_iu_0.$$
The hydrodynamic reductions of (\ref{3Dgen1}) is given by the  pair of semi-hamiltonian (1+1)-dimensional systems
$$r^i_y=e^{u_0}\Big(1-\frac{1}{q_i}\Big)r^i_x, \qquad r^i_{t}=e^{u_0+u_1}\Big(\frac{1}{(p_i-1)q_i}+1\Big)r^i_x.$$

Chain (\ref{3Dgen1}) is the first member of an infinite hierarchy of pairwise commuting flows where the corresponding "times" are $t_1=t,\, t_2,\, t_3,...$. These flows and their hydrodynamic reductions can be described in terms of the generating function $U(z)=u_1+u_2z+u_3z^2+...$  The hierarchy is given by 
$$D(z)u_0=e^{U(z)}\Big(u_{0,y}+U(z)_y-e^{u_0}U(z)_x\Big),$$ 
$$D(z_1)U(z_2)=e^{u_0+U(z_1)}U(z_2)_x+(1+z_1)e^{U(z_1)}\Big(e^{u_0}\frac{U(z_1)_x-U(z_2)_x}{z_1-z_2}-\frac{U(z_1)_y-U(z_2)_y}{z_1-z_2}\Big),$$
where $D(z)=\frac{\partial}{\partial t_1}+z \frac{\partial}{\partial t_2}+z^2 \frac{\partial}{\partial t_3}+...$ The reductions can be written as 
$$D(z)r^i=e^{u_0+U(z)}\Big(1+\frac{1+z}{(p_i-1-z)q_i}\Big)r^i_x.$$

Other (2+1)-dimensional integrable chains related to 2-dimensional vector spaces of solutions for system (\ref{sys}) 
are degenarations of (\ref{3Dgen1}). In particular $F=\gamma_1 e^{u_1}p+\gamma_2 (p+u_2)$ leads to the following 
(2+1)-dimensional integrable generalization of (\ref{3Dgen}):
$$
u_{k,t}= e^{u_1}u_{k+1,x}+ u_{k+1,y}+u_2 u_{k,y}, \qquad k=1,2,3,....
$$

{\bf Conjecture.} Any chain of the form (\ref{3D}) integrable by the hydrodynamic reduction method 
is a degeneration of (\ref{3Dgen1}).

We are planning to consider the problem of classification  of integrable chains (\ref{3D}) in a separate paper.

\section{Infinitesimal symmetries of triangular GT-systems}

A scientific way to construct the functions $g_3,g_4,...$ for different cases from Proposition 1 is related to infinitesimal symmetries of the corresponding GT-system\footnote{Note that these functions are not unique because of the triangular group of symmetries (\ref{tran}) acting on the fields $u_3,u_4,...$ }. The whole Lie algebra of symmetries is one the most important algebraic structures related to any triangular GT-system (\ref{gttr}). In particular, this algebra acts on the  hierarchy of the commuting flows for the corresponding chain (\ref{genchain}).

A vector field
\begin{equation} \label{SS}
S= \sum_{j=1}^N X(p_{j},u_{1},...,u_{s}) \frac{\partial}{\partial p_j}+\sum_{m=1}^{\infty}
Y_{m}(u_{1},...,u_{k_{m}}) \frac{\partial}{\partial u_m}, \qquad \frac{\partial Y_{m}}{\partial u_{k_{m}}}
\ne 0
 \end{equation}
is called a {\it symmetry} of the triangular GT-system (\ref{gttr}) if it
commutes with all $\partial_{i}.$ Notice that it follows from the definition
that
$$
S (\partial_{i} u_{1})= \partial_{i}(Y_{1}).
$$
We call (\ref{SS}) a symmetry of shift $d$ if $k_{m}=m+d$ for
$m>>0.$ Let $M$ be the minimal integer such that $k_{m}=m+d, m>M.$
If the functions $g_{i}, i=1,...,M+d$  from
(\ref{gttr}) are known, then the functions $X, Y_{1},...Y_{M}$
can be found from the compatibility conditions
$$
S (\partial_{i} p_{j})=\partial_{i} S(p_{j}), \qquad S (\partial_{i} u_{k})=\partial_{i}
S(u_{k}), \quad k=1,...,M.
$$
The functions  $Y_{M+1}, Y_{M+2},...$ can be chosen arbitrarily. After that $g_{M+d+1}, g_{M+d+2},... $ are uniquely defined by the remaining compatibility conditions.

{\bf The generic case 1, $\bf a_1$}.  Looking for
symmetries of shift one, we find $X=Y_{1}=0$ and $M=1$. Hence
without loss of generality we can take
$$
S=\sum_{m=2}^{\infty} u_{m+1} \frac{\partial}{\partial u_m}
$$
for the symmetry. This fact gives us a way to construct all
functions $g_{i}, i>3$ in the infinite triangular extension for the
case 1, $a_{1}.$ Indeed, it follows from the commutativity
conditions  $S (\partial_{i} u_{k})=\partial_{i} S(u_{k})$ that
$g_{k+1}=S(g_{k}),$ where $k=2,3,...$. In particular,
$$
g_{3}=\frac{(p_{j}-u_{1})(2 p_{j} u_{2}-p_{j}-u_{2}^{2}) u_{3}}{u_{1}(u_{1}-1)
(p_{j}-u_{2})^{2}}.
$$
The functions $g_i$ thus constructed are not linear in $u_3.$
The corresponding chain  (\ref{genchain}) is equivalent to the chain constructed in Section 5 but not so simple.

It would be interesting to describe the Lie algebra of all symmetries in this case. Here we present the essential part for symmetry of shift 2:
$$
X=\frac{p_{j} (p_{j}-1) u_{3}^{2}}{(p_{j}-u_{2}) u_{2} (u_{2}-1)}, \qquad
Y_{1}=\frac{u_{1} (u_{1}-1) u_{3}^{2}}{(u_{1}-u_{2}) u_{2} (u_{2}-1)},
$$
$$
Y_{2}=-\frac{3}{2} u_{4}+\frac{ (2 u_{1}-1) u_{3}^{2}}{  u_{2}
(u_{2}-1)}+u_{3}. \qquad \square
$$

{\bf The case 1, $\bf d_1$}.  One can add fields $u_3,...$ in such a way that the whole triangular GT-system admits the
following symmetry of shift 1:

$$S=\frac{u_2}{u_1(u_1-1)}\sum_{i=1}^Np_i(p_i-1)\frac{\partial}{\partial p_i}+\sum_{i=1}^{\infty}u_{i+1}\frac{\partial}{\partial u_i}.$$
As in the previous example, one can easily recover the whole GT-system.
For example,
$$\partial_iu_3=\left(\frac{u_3(p_i+u_1-1)}{u_1(u_1-1)}+\frac{2u_2^2p_i(p_i-1)}{u_1^2(u_1-1)^2}\right)\partial_iu_1. \qquad \square$$

Below we describe the symmetry algebra for the  case 5, $c_2$ (in particular, for the Benney chain).

{\bf The case 5, $\bf c_2$}.  For the triangular GT-system (\ref{gtBenni}), (\ref{gtBenni1}) there exists an infinite Lie algebra of symmetries $S_i, i\in \Z,$ where $S_i$ is a symmetry of shift $i$. The simplest symmetries are the following:
$$
S_{-2}=  \frac{\partial}{\partial u_1}+\sum_{i=3}^\infty \Big( -u_{i-2}+\sum_{k+m=i-3}
 u_k u_{m}-\sum_{k+m+l=i-4}
 u_k u_{m} u_{l}+\cdots\Big) \frac{\partial}{\partial u_i},
$$
$$
S_{-1}=\sum_{j=1}^N   \frac{\partial}{\partial p_j}+ \sum_{i=1}^\infty (i-1) u_{i-1} \frac{\partial}{\partial u_i},
$$
$$
S_{0}=\sum_{j=1}^N p_j  \frac{\partial}{\partial p_j}+ \sum_{i=1}^\infty (i+1) u_{i} \frac{\partial}{\partial u_i},
$$
$$
S_1= \sum_{j=1}^N (p_{j}^2+3 u_1) \frac{\partial}{\partial p_j}+
\sum_{i=1}^\infty (i+3) u_{i+1}\frac{\partial}{\partial u_i}+
\sum_{i=2}^\infty\sum_{k+m=i} u_k u_{m} \frac{\partial}{\partial
u_i}+\sum_{i=2}^\infty3(i-1)u_1u_{i-1}\frac{\partial}{\partial u_i},
$$

$$
S_2= \sum_{j=1}^N (p_{j}^3+4 u_1 p_j+5 u_2) \frac{\partial}{\partial
p_j}+\sum_{i=1}^\infty  (i+5) u_{i+2}\frac{\partial}{\partial
u_i}+\sum_{i=1}^\infty 4iu_1u_i\frac{\partial}{\partial
u_i}+\sum_{i=2}^\infty 5(i-1)u_2u_{i-1}\frac{\partial}{\partial
u_i}+$$
$$\sum_{i=1}^\infty\sum_{k+m=i+1} 3 u_k
u_{m}\frac{\partial}{\partial u_i}+\sum_{i=3}^\infty\sum_{k+m+l=i}
u_k u_{m} u_{l} \frac{\partial}{\partial u_i}.
$$
The whole algebra is generated by $S_1,S_2,S_{-1},S_{-2}.$
It is isomorphic to the Virasoro algebra with zero central charge.

Let $D_{t_{i}}$ be the vector fields corresponding
to commuting flows for the Benney chain. Here $D_{t_{1}}=D_x, \, D_{t_{2}}=D_t$. Then
the commutator relations
$$
[S_1, D_{t_{i}}]=(i+1) D_{t_{i+1}}
$$
hold. Thus the vector field $S_1$ plays the role of a master-symmetry for the Benney hierarchy.
$\square$

{\bf The case 6}. In this case there exist infinitesimal symmetries of form
$$T_i=u_{i+1}\frac{\partial}{\partial u_1}+u_{i+2}\frac{\partial}{\partial u_2}+...,~i=0,1,2,...$$
$$S_i=\sum_{j=1}^{N}p_j^{i+1}\frac{\partial}{\partial p_j}+u_{i+2}\frac{\partial}{\partial u_2}+2u_{i+3}\frac{\partial}{\partial u_3}+3u_{i+4}\frac{\partial}{\partial u_4}+...,~i=-1,0,1,2,...$$
Note that $[S_i,S_j]=(j-i)S_{i+j},~[T_i,T_j]=0,~[S_i,T_j]=jT_{i+j}$.   $\square$

\end{document}